\documentclass[a4paper,superscriptaddress,aps,prb,twocolumn,floatfix,citeautoscript]{revtex4}

\newcommand{\vect}[1]{\mathbf{#1}}
\usepackage{graphicx}
\usepackage{color}
\usepackage{amsmath}
\usepackage{amssymb}
\usepackage[english]{babel}
\usepackage[utf8]{inputenc}

\usepackage{bm}
\usepackage{url}
\usepackage[colorlinks=true,linkcolor=blue,citecolor=blue,urlcolor=blue]{hyperref}
\graphicspath{{figures/}}

\begin{document}
%
%
\title{Many-Body Effective Energy Theory: photoemission at strong correlation}
\newcommand{\lpt}{Laboratoire de Physique Th\'eorique, Universit\'e de Toulouse, CNRS, UPS, France}
\newcommand{\lcpq}{Laboratoire de Chimie et Physique Quantiques, Universit\'e de Toulouse, CNRS, UPS, France}
\newcommand{\etsf}{European Theoretical Spectroscopy Facility (ETSF)}
\affiliation{\lpt}
\affiliation{\lcpq}
%
%
\affiliation{\etsf}
\author{S. Di Sabatino}
\affiliation{\lpt}
\affiliation{\etsf}
\author{J.A. Berger}
\affiliation{\lcpq}
\affiliation{\etsf}
\author{P. Romaniello}
\affiliation{\lpt}
\affiliation{\etsf}
%

\begin{abstract}
In this work we explore the performance of a recently derived many-body effective energy theory for the calculation of photoemission spectra in the regime of strong electron correlation. We apply the theory to paramagnetic MnO, FeO, CoO, and NiO, which are typical examples of strongly correlated materials and, therefore, a challenge for standard theories. We show that our methods open a correlation gap in all the oxides studied without breaking the symmetry. Although the materials seem similar, we show that an analysis of the occupation numbers reveals that the nature of the gap is not the same for these materials. Overall the results are very promising, although improvements are clearly required, since the band gap is overestimated for all the systems studied. We indicate some possible strategies to further develop the theory.

%
\end{abstract}
\date{\today}
\maketitle
\section{Introduction}
One of the grand challenges of materials science and condensed matter physics today is the development of predictive
and reliable approaches to describe and
understand materials and, ultimately, to predict
new ones. As a general rule, a common theme
for most of the exciting new materials
discovered recently is the presence of
strong electronic correlations.  Strong electron correlation can be nicely illustrated and understood with the simple case of the $H_2$ molecule at dissociation: the two (antiparallel) electrons in the system localize each on one site with equal probability. In such a scenario, the (singlet) wave function of the system cannot be accurately described by a single Slater determinant, and a mean-field description would simply fail. This is not just an academic example: this kind of scenario is ubiquitous in strongly correlated materials such as NiO and, more generally, in transition metal oxides.\cite{stefano} These systems exhibit remarkable electronic and magnetic properties, such as metal-insulator transitions, half-metallicity, or unconventional superconductivity, which make them among the most attractive and versatile materials with direct applications in various technological fields from nonlinear optics to sensors and
catalysis. The peculiar properties of these materials originate from their incompletely filled d- or f -electron shells with narrow energy bands, which require a particularly accurate theoretical treatment of electron correlation. 

A unique
source of precious information about electronic structure and excitations in materials is photoemission. The interpretation of
the experimental data is far from obvious, due to the coexistence and interplay of various physical
mechanisms underlying the observed spectral features. Theory hence represents an essential
complementary tool for the analysis of the experiments as well as prediction of material properties.
The correct description of the electronic structure is, moreover, essential for modeling more advanced
experiments that involve pump-and-probe techniques,\cite{Pogna} high-intensity ultra-short pulses (FEL
sources),\cite{Pompili} and novel experiments with unprecedented high temporal and spatial resolution.\cite{Man}

There are various ways to tackle the problem. One could use, for example, a model Hamiltonian devised ad hoc for this kind of systems. However in this case one should rely on particular parametrizations of the Hamiltonian or of the electron-electron interaction, which makes the theory non predictive. Therefore it would be desirable to describe these systems and their physics from first principles. In this context Many-Body Perturbation Theory, within the so-called $GW$ approximation to electron correlation, is the method of choice for calculations of photoemission spectra of many materials.  However, $GW$  suffers from some fundamental shortcomings, and, in particular, it does not capture strong correlation, unless one treats the system in a magnetically ordered phase. A deep problem is, indeed, the description of the paramagnetic phase. An alternative approach based on Green's functions that can treat strongly correlated systems is dynamical mean-field theory (DMFT).\cite{Georges} However, efforts to make DMFT a first-principles method are still ongoing.\cite{Biermann} 
Therefore, there is an increasing effort to explore novel routes to calculate photoemission spectra (PES) accurately.\cite{lani,stefano} 
Very recently an exact expression for the interacting spectral function in terms of the Kohn-Sham one of DFT has been derived and successfully applied to model systems;\cite{Kurth} its feasibility for realistic systems is still to be explored. Promising results have been reported for solids \cite{sharma_PRB08,sharma_JCTC} using reduced density-matrix functional theory (RDMFT).\cite{Gilbert} Similarly to DFT, the RDMFT framework allows for the calculation of all the ground-state expectation values as functionals of the one-body reduced density matrix, provided that the functional is known. This, however, is in general not the case. In particular, for spectral functions approximations have to be used.\cite{PhysRevLett.110.116403,stefanoJCP} 

We have recently derived a new method, the Many-body Effective Energy Theory (MEET), for the calculation of photoemission spectra in terms of reduced density matrices.\cite{stefano} Simple approximations in terms of  low-order density matrices give accurate spectra in model systems in the weak as well as strong correlation regime. Preliminary results on NiO are promising. In particular, our method correctly predicts paramagnetic NiO to be an insulator without breaking the symmetry, contrary to standard approaches. In this work we provide further evidences of the capabilities of the method, but also of its drawbacks and possible solutions. The paper is organized as follows. In Sec.\ \ref{Sec:Theory} we will report the key equations of the many-body effective energy theory. We provide also the basic equations of RDMFT, since we use this framework for practical calculations. In Sec.\ \ref{Sec:Results} the spectral function of the bulk MnO, FeO, CoO, and NiO in the paramagnetic phase are compared to available experimental data and analyzed in terms of occupation numbers and atomic orbital character. Conclusions and outlooks are given in Sec.\ \ref{Sec:Conclusions}.
\section{Theoretical framework\label{Sec:Theory}}
In this section we give the key equations of the many-body effective energy theory derived in Ref.\ [\onlinecite{stefano}] as well as the basic equations of reduced density-matrix functional theory.
\subsection{The Many-Body Effective Energy Theory}
The spectral function, which is related to photoemission spectra, can be expressed in terms of the imaginary part of the one-body
Green's function $G$ as 
$A(\omega)=|\Im  G (\omega)|/\pi$. The Many-Body Effective Energy Theory expresses $A(\omega)$ in terms of $p$-body density matrices
 \begin{multline*}
\Gamma^{(p)}(\mathbf{x}_1...\mathbf{x}_p,\mathbf{x}_1'...\mathbf{x}_p')=\\
\frac{N!}{(N-p)!}\int d \mathbf{x}_{p+1}...d\mathbf{x}_N\Psi^*(\mathbf{x}_1'...\mathbf{x}'_p,\mathbf{x}_{p+1}...\mathbf{x}_N)\\
\times\Psi(\mathbf{x}_1...\mathbf{x}_p,\mathbf{x}_{p+1}...\mathbf{x}_N).
\end{multline*} 
To achieve this we start from the spectral representation of $G$ at zero temperature and we concentrate on the diagonal elements of $G$, which are related to photoemission spectra. We follow three main steps:
\begin{enumerate}
\item
Treat separately the part of $G$  related to direct photoemission ($G_{ii}^{R}$) and inverse photoemission ($G_{ii}^{A}$) as it is done in experiments:
\begin{eqnarray}
	G_{ii}(\omega)&=&G_{ii}^{R}(\omega)+G_{ii}^{A}(\omega)\nonumber\\
	&=&\sum_k\frac{B_{ii}^{k,{R}}}{\omega-\epsilon_k^{{R}}}
	+\sum_k\frac{B_{ii}^{k,{A}}}{\omega-\epsilon_k^{{A}}},
	\label{eqn:CapEET_SR_G}
	\end{eqnarray}
	where $\epsilon_k^{R}=E_0-E_k^{N-1}$ and $\epsilon_k^{A}=E_k^{N+1}-E_0$ are energies measured in direct and inverse photoemission, respectively, $B_{ii}^{k,{R}}=\langle\Psi_0|\hat{c}_j^\dagger|\Psi_k^{N-1}\rangle\langle\Psi_k^{N-1}|\hat{c}_i|\Psi_0\rangle$, $B_{ii}^{k,{A}}=\langle\Psi_0|\hat{c}_i|\Psi_k^{N+1}\rangle\langle\Psi_k^{N+1}|\hat{c}_j^\dagger|\Psi_0\rangle$, with $E_0$ and $\Psi_0$ the ground-state energy and wave function
of the $N$-electron system and $E^{N\pm1}_
k$ and $\Psi^{N\pm1}_
k$ the $k$th state
energy and wave function of the ($N \pm 1$)-electron system.
	
 \item
  For each part of $G$, introduce a dynamical effective energy $\delta^{R/A}(\omega)$ that describes all its poles (see Fig.~\ref{figure3}). This effective energy has a well-defined form, which can be expressed in terms of $p$-body density matrices.

\begin{figure}[t]
\begin{center}
\includegraphics[width=0.2\textwidth]{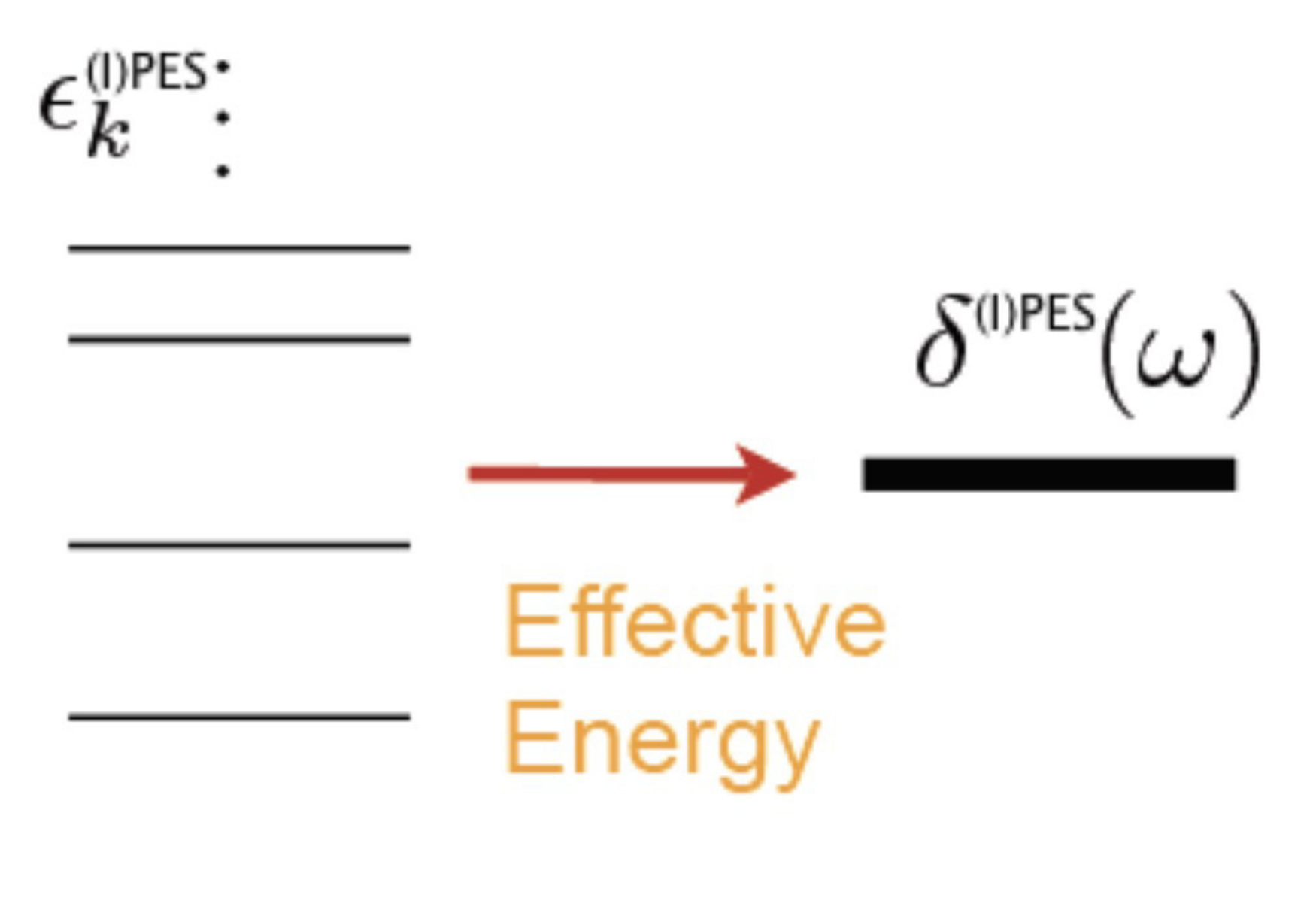}
\caption{A pictorial representation of the many-body effective-energy theory: the frequency-dependent effective energy $\delta^{R/A}_{i}(\omega)$ is introduced to account for all the poles $\epsilon^{E/A}_k$ of $G$.}
\label{figure3}
\end{center}
\end{figure}

 \item
Truncate the series in terms of density matrices to low order to obtain approximations that can be used in practice.
\end{enumerate}
We choose to work in the basis of natural orbitals $\phi_{i}$, \textit{i.e.}, the orbitals which diagonalize the 1-RDM, $\Gamma^{(1)}(\vect{x},\vect{x}')=\sum_{i} n_{i}\phi_{i}(\vect{x})\phi^*_{i}(\vect{x}')$, where $n_{i}$ are the occupation numbers. The final result should not depend on the choice of the basis set; however, since in practice we truncate the series, it does. We found that the basis of natural orbitals is the most appropriate choice. In this basis $\sum_k B_{ii}^{k,R}=n_{i}$ and $\sum_k B_{ii}^{k,A}=(1-n_{i})$. We hence arrive at:
\begin{eqnarray}
G_{ii}^{R}(\omega)&=&\sum_k\frac{B_{ii}^{k,{R}}}{\omega-\epsilon_k^{{R}}}=\frac{n_{i}}{\omega-\delta^R_{i}(\omega)},
\label{eqn:CapEET_EET_G} \\
G_{ii}^{A}(\omega)&=&\sum_k\frac{B_{ii}^{k,{A}}}{\omega-\epsilon_k^{{A}}}=\frac{1-n_{i}}{\omega-\delta^A_{i}(\omega)},
\label{eqn:CapEET_EET_G_2} 
\end{eqnarray}
(which is possible since $\delta^{R/A}_{i}(\omega)$ does not depend on the index $k$), from which, using a common denominator, we get  
\begin{eqnarray}
 \delta^R_{i}(\omega)&=&\frac{1}{G^R_{ii}(\omega)}\sum_k\frac{\langle\Psi_0|\hat{c}_{i}^\dagger|\Psi_k^{N-1}\rangle\langle \Psi_k^{N-1}|[\hat{c}_{i},\hat{H}]|\Psi_0\rangle}{\omega-\epsilon_k^{R}}\nonumber\\&=&\frac{\tilde{G}_{ii}^R(\omega)}{G^R_{ii}(\omega)},
 \label{eqn:CapEET_EET_2_delta}\\
  \delta^A_{i}(\omega)&=&\frac{1}{G^A_{ii}(\omega)}\sum_k\frac{\langle\Psi_0|[\hat{c}_{i},\hat{H}]|\Psi_k^{N+1}\rangle \langle \Psi_k^{N+1}|\hat{c}^\dagger_{i}|\Psi_0\rangle}{\omega-\epsilon_k^{A}}\nonumber\\
  &=&\frac{\tilde{G}_{ii}^A(\omega)}{G^A_{ii}(\omega)}.
 \label{eqn:CapEET_EET_A_delta}
\end{eqnarray}
Note that $\delta_{i}^R\neq\delta_{i}^A$, so that we have an effective energy for the removal part and an effective energy for the addition part. In principle one can follow a similar procedure also for the removal and addition poles in \eqref{eqn:CapEET_EET_2_delta} and \eqref{eqn:CapEET_EET_A_delta}, respectively, by introducing other two effective energies that can account for all the poles. The process could be continued over and over again.  However,  one wishes  to  truncate  the  expression  for $\delta_i^{R/A}$ since, in  practice, one would like to use simple expressions. There are several way to truncate. In Ref. [\onlinecite{stefano}] we chose a truncation that guarantees the exact results for the Hubbard dimer at one-half filling at all orders. This is obtained by assuming that at a certain order the poles of $G^{R/A}_{ii}$, $\tilde{G}^{R/A}_{ii}$,..., expressed in terms of the respective effective energies $\delta^{R/A}_i$, $\tilde{\delta}^{R/A}_i$,..., are the same. This choice was motivated by the fact that the physics underlying the atomic limit of the Hubbard dimer is common also to realistic strongly correlated materials, which are our target. The obtained expressions contain commutators of the creation and annhilation operators with the Hamiltonian of the systems, which can be worked out, and expressed in terms of n-body density matrices. For example, the first approximation for $\delta^{R/A}_i(\omega)$ can be expressed in terms of one- and two-body density matrices as
\begin{widetext}
\begin{align}
 \delta_i^{R,(1)}&=\frac{\tilde{n}_i^R}{n_i}=h_{ii}+\frac{1}{n_i}\sum_{jkl}V_{ijkl}\Gamma^{(2)}_{klji},\label{eqn:ChapEET_deltaRGamma}\\
 \delta_i^{A,(1)}&=\frac{\tilde{n}_i^A}{1-n_i}=h_{ii}+\frac{1}{1-n_i}
 \left[\sum_{j}(V_{ijij}-V_{ijji})n_{j}-\sum_{jkl}V_{ijkl}
  \Gamma_{klji}^{(2)}\right].
  \label{eqn:ChapEET_deltaAGamma}
\end{align}
\end{widetext}
with $\tilde{n}_i^R=\langle\Psi_0|\hat{c}_{i}^\dagger|[\hat{c}_{i},\hat{H}]|\Psi_0\rangle$ and $\tilde{n}_i^A=\langle\Psi_0|[\hat{c}_{i},\hat{H}]\hat{c}_{i}^\dagger|\Psi_0\rangle$. 
 Here $h_{ij}=\int d\vect{x}\phi_i^*(\vect{x})h(\vect{r})\phi_j(\vect{x})$ are the matrix elements of the one-particle noninteracting Hamiltonian $h(\vect{r})=-\nabla^2/2+v_{\text{ext}}(\vect{r})$, and 
$$V_{ijkl}=\int d\vect{x}d\vect{x}'\phi^*_i(\vect{x})\phi^*_j(\vect{x}')v_c(\vect{r},\vect{r}')\phi_k(\vect{x})\phi_l(\vect{x}'),$$
are the matrix elements of the Coulomb interaction $v_c$. 

\begin{figure}[h]
\begin{center}
\includegraphics[width=0.4\textwidth]{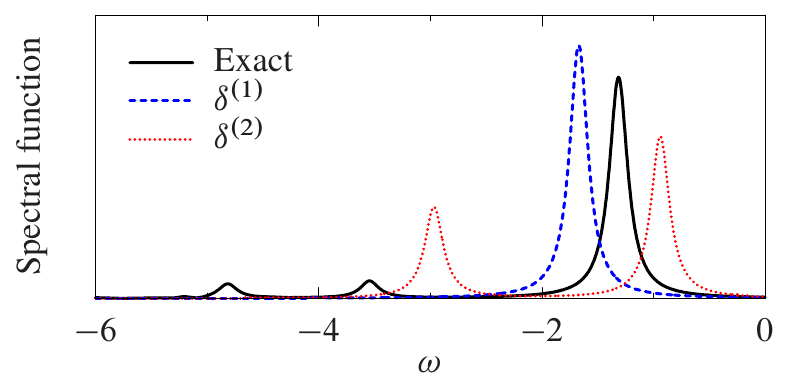}
\caption{A pictorial representation of the effect of approximations to $\delta_i(\omega)$ on a toy spectrum (removal part of a one component of $G$): exact (black solid line) \textit{vs} MEET at $\delta^{(1)}$ (blue dashed line) and $\delta^{(2)}(\omega)$ (red dotted line) levels.}
\label{Fig:toy-spectrum}
\end{center}
\end{figure}

As commented in Ref. [\onlinecite{stefano}] the various approximations $\delta^{R/A,{n}}_i(\omega)$ are related to the first $n$-th moments $\mu^{R/A}_{n,i}=\sum_kB^{k,R}_{ii}(\epsilon_k^R)^n /\sum_k B^{k,R}_{ii}$ of the $G^{R/A}_{ii}(\omega)$. This allows for a more compact expression of $G^{R/A}_{ii}(\omega)$ as a continued fraction of moments
\begin{equation}
G^{R}_{ii}=\frac{n_i}{\omega-\mu^R_{1,i}\frac{\omega-\mu^R_{1,i}...}{\omega-\frac{\mu^R_{2,i}}{\mu^R_{1,i}}...}},
\end{equation}
(and similarly for $G^{A}_{ii}$). At the level of $\delta^{R/A,(1)}$, the Green's function depends only on the first moment, while neglecting all the higher-order frequency-dependent corrections. As illustrated in Fig. (\ref{Fig:toy-spectrum}) this means that each component $G^{R/A}_{ii}$ shows only one pole which is a weighted average of all the poles of $G^{R/A}_{ii}$. If each component of $G$ has a predominant quasiparticle peak, this is a good approximation, provided that the approximation to the first moment is accurate enough. At the level of $\delta^{R/A,(2)}$  the Green's function depends on the first and second moments; since now the corrections are frequency-dependent more poles appear (namely two removal and two addition poles for each component of $G$, which are visible if the corresponding weights are nonzero). This approximation tends to reproduce the two most dominant removal/addition peaks for each component of $G$, as shown in Fig. (\ref{Fig:toy-spectrum}). Higher-order moments will produce more poles; however approximations become quickly uncontrolled, which can lead to unphysical results.

Using Eqs \eqref{eqn:CapEET_EET_G} and \eqref{eqn:CapEET_EET_G_2}, the spectral function can then be written as
\begin{equation}
A_{ii}(\omega)=n_{i}\delta(\omega-\delta^R_{i}(\omega))+(1-n_{i})\delta(\omega-\delta^A_{i}(\omega)),
\label{eqn:ChapEET_SF}
\end{equation}
which satisfies the well-known sum rule $\int^{\infty}_{-\infty}A_{ii}(\omega)d\omega=1$. 

Our method is general, and can be used
together with any theoretical approach from which the $p$-body density matrices can be obtained. For practical calculations we restrict here to the first approximations to $\delta_i(\omega)^{R/A}$, given in Eqs (\ref{eqn:ChapEET_deltaAGamma}), and we use reduced density matrix functional theory to obtain the one- and two-body reduced density matrices. 

More details about the MEET can be found in Ref.\ [\onlinecite{stefano}]. 
\subsection{Reduced density matrix functional theory}
Within RDMFT the ground-state properties of a physical system are functionals of the ground-state density matrix,\cite{PhysRev.97.1474,Gilbert75} since there exists a one-to-one mapping between the (non-degenerate) ground-state wavefunction of the system and the corresponding density matrix.\cite{Gilbert75} In particular the ground-state total energy is a functional of the one-body reduced density matrix $\Gamma^{(1)}$ (1-RDM)  and it can be written as  
\begin{equation}
E_{\text{tot}}[\Gamma^{(1)}]=E_{\text{kin}}[\Gamma^{(1)}]+E_{\text{ext}}[\Gamma^{(1)}] + E_{\text{Hxc}}[\Gamma^{(1)}], 
\end{equation}
where $E_{\text{kin}}$, $E_{\text{ext}}$, and $E_{\text{Hxc}}$, are the kinetic energy, the energy due to the coupling to an external potential, and the Hartree and exchange-correlation energies, respectively. Energy minimization under the constraint that $\Gamma^{(1)}$ is (ensemble) $N$-representable (which corresponds to the natural occupation numbers being $0\leq n_i\leq 1$ with $\sum_in_i=N$ ), determines the exact 1-RDM. In practice, however, approximations to the exchange-correlation energy
\begin{equation}
E_{\text{xc}}[\Gamma^{(1)}]=\frac{1}{2}\int \! d \mathbf{x}\, d \mathbf{x}^\prime \, v_{\text{c}}(\mathbf{x},\mathbf{x}^\prime)
\Gamma^{(2)}_{\text{xc}}[\Gamma^{(1)}](\mathbf{x},\mathbf{x}^\prime ; \mathbf{x},\mathbf{x}^\prime) 
\end{equation}
where $\Gamma^{(2)}_{\text{xc}}[\Gamma^{(1)}](\mathbf{x},\mathbf{x}^\prime ; \mathbf{x},\mathbf{x}^\prime) =\Gamma^{(2)}[\Gamma^{(1)}](\mathbf{x},\mathbf{x}^\prime ; \mathbf{x},\mathbf{x}^\prime)-\Gamma^{(1)}(\mathbf{x},\mathbf{x})\Gamma^{(1)}(\mathbf{x}',\mathbf{x}')$ is the xc contribution to the 2-body reduced density matrix (2-RDM), are needed. Several approximations have been proposed and most of them are implicit functionals of the 1-RDM; they are explicit functionals of the natural orbitals and occupation numbers.\cite{PhysRev.97.1474}  
In this work we  use the so-called power functional \cite{sharma_PRB08} to approximate $\Gamma^{(2)}_{\text{xc}}$ as
\begin{equation}
\Gamma^{(2)}_{\text{xc}}[\Gamma^{(1)}](\mathbf{x},\mathbf{x}^\prime ; \mathbf{x},\mathbf{x}^\prime)= -\Gamma^{(1)\alpha}(\mathbf{x},\mathbf{x}^\prime)\Gamma^{(1)\alpha}(\mathbf{x}^\prime, \mathbf{x}),
\label{Eqn:POWER}
\end{equation}
where $\Gamma^{(1)\alpha}(\mathbf{x},\mathbf{x}^\prime)=\sum_{j}^{}n^\alpha_{j}\phi_{j}(\mathbf{x})\phi^*_j(\mathbf{x}^\prime)$ and $0.5\leq\alpha\leq1$.
This factorization for $\Gamma^{(2)}$ is used both in the total energy and in $\delta^{R/A,(1)}_{i}$. After a total energy minimization the optimal natural orbitals and occupation numbers are used to calculate $\delta^{R/A,(1)}_{i}$.
\section{Results and Discussion\label{Sec:Results}}
In this section we report the results for the spectral function of the bulk transition metal (TM) oxides MnO, FeO, CoO, and NiO in the paramagnetic (PM) phase. 
We implemented our method in the open-source full-potential linearized augmented plane wave (FP-LAPW) code ELK,\cite{elk} with practical details of the calculations
following the scheme described in Ref.~[\onlinecite{sharma_PRB08}]. 

The oxides under study, which crystallize in rocksalt structure in
the paramagnetic phase, show antiferromagnetic (AF)
behavior below their respective N\'eel temperatures. For the PM phases we use half of the experimental lattice constant of the corresponding AF phases, which are 8.863 \AA\, for AF MnO, 8.666 \AA\, for AF FeO, 8.499 \AA\, for AF CoO, and
8.341 \AA\, for AF NiO. For all systems we used a $8\times8\times8$ k-point grid and we fixed the parameter $\alpha$ of the power functional at 0.65, as suggested in literature at least for NiO and MnO.\cite{sharma_PRB08,sharma_JCTC}

Both GGA and LSDA calculations yield a too small gap compared to experiments in the case of the AF MnO and NiO, while AF CoO and FeO are predicted to be metallic.\cite{Roedl} This is caused by the inappropriateness of the KS approach, which is
based on a semilocal treatment of the XC. R\"{o}dl \textit{et al.} \cite{Roedl} showed that a perturbative $GW$ approach on top of a KS starting point obtained using the nonlocal screened hybrid functional HSE03 can well describe the AF phase of these systems. The description of the PM phase, instead, remains still a challenge for ab initio methods. Here we show that our approach can open a gap in the PM phase. To treat the PM phase we model the paramagnetic oxides as nonmagnetic. Although this might be not the most appropriate way to model the PM phase (see, e.g., Ref. [\onlinecite{zunger}]), it is a standard practice in electronic structure calculations.

\begin{figure*}[t]
\begin{center}
\includegraphics[width=0.45\textwidth]{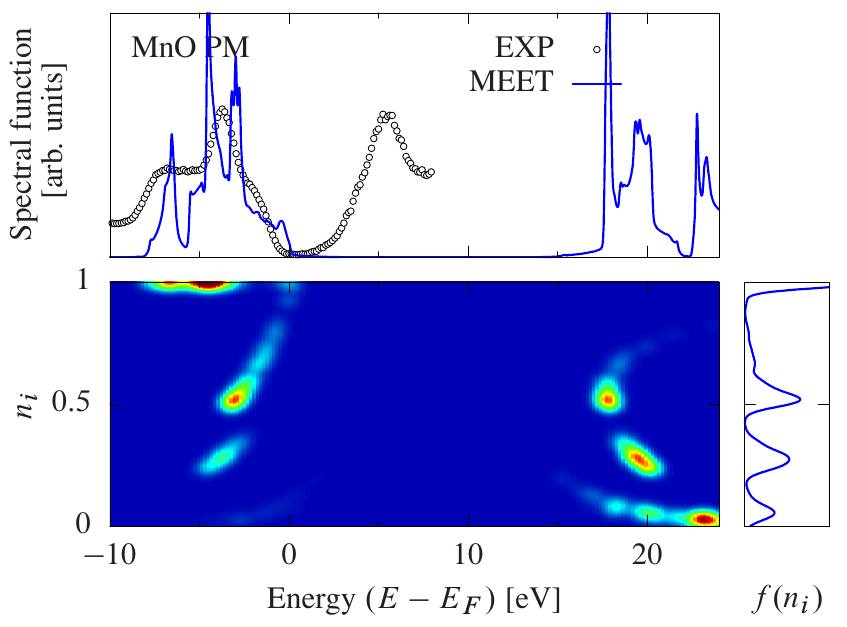}
\includegraphics[width=0.45\textwidth]{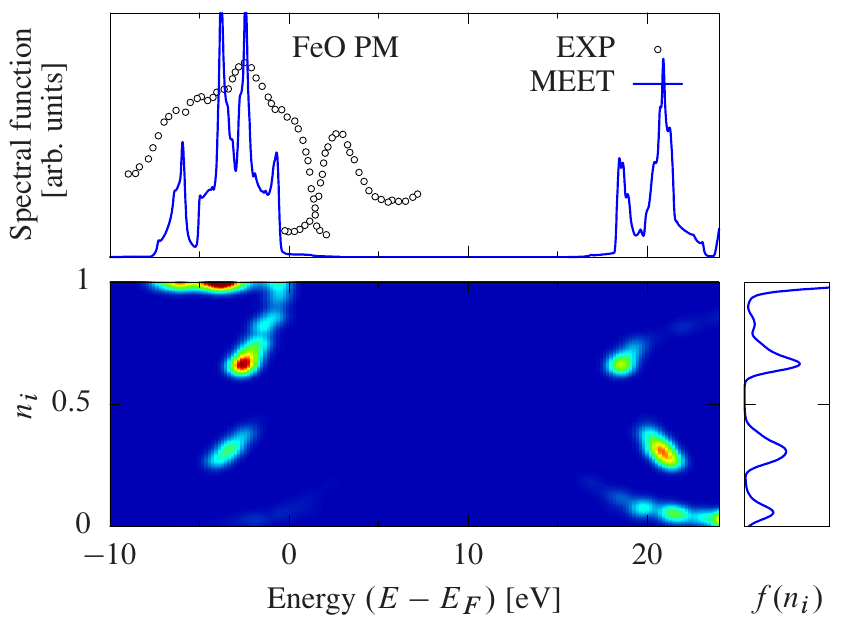}\\
\vspace{10pt}
\includegraphics[width=0.45\textwidth]{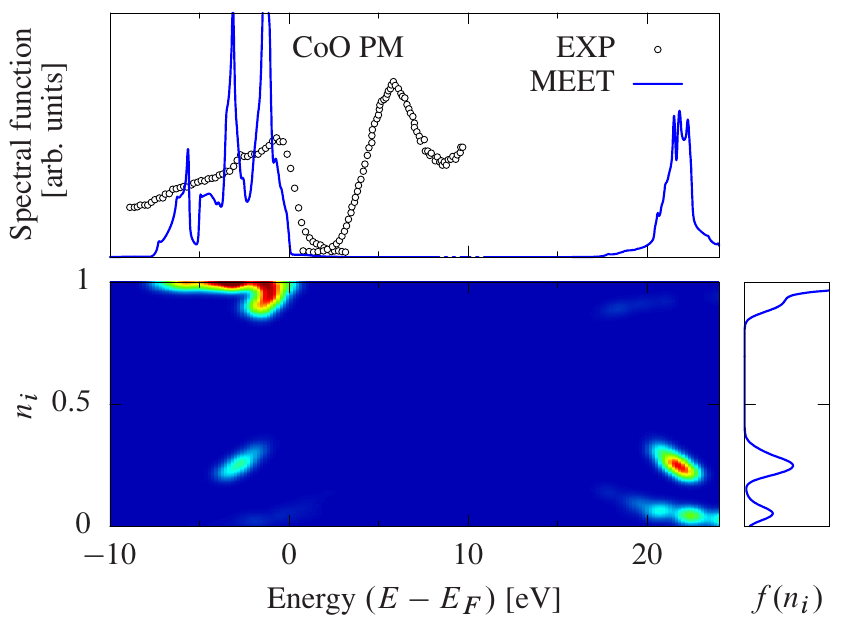}
\includegraphics[width=0.45\textwidth]{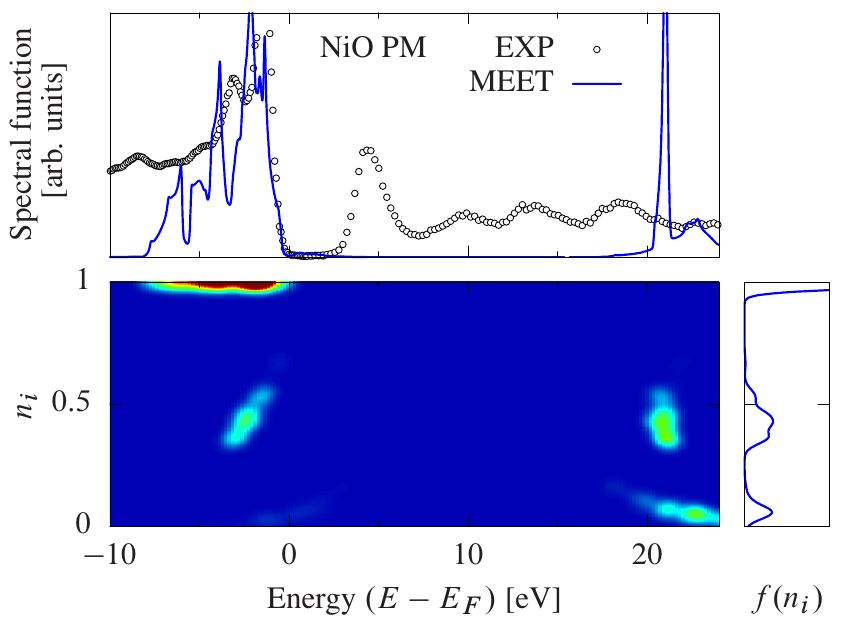}
\caption{Paramagnetic phase of bulk MnO, FeO, CoO and NiO: experimental photoemission spectrum \cite{PhysRevB.44.1530,PhysRevLett.39.1229,PhysRevLett.53.2339,PhysRevB.54.10245,PhysRevB.44.6090}  vs MEET spectrum ($\alpha=0.65$). The color map and the distribution $f(n_i)$ illustrate the occupation numbers $n_i$ that play a role into the spectrum for the reported energy range.}
\label{figurePM}
\end{center}
\end{figure*}

\begin{figure}[t]
\begin{center}
\includegraphics[width=0.48\textwidth]{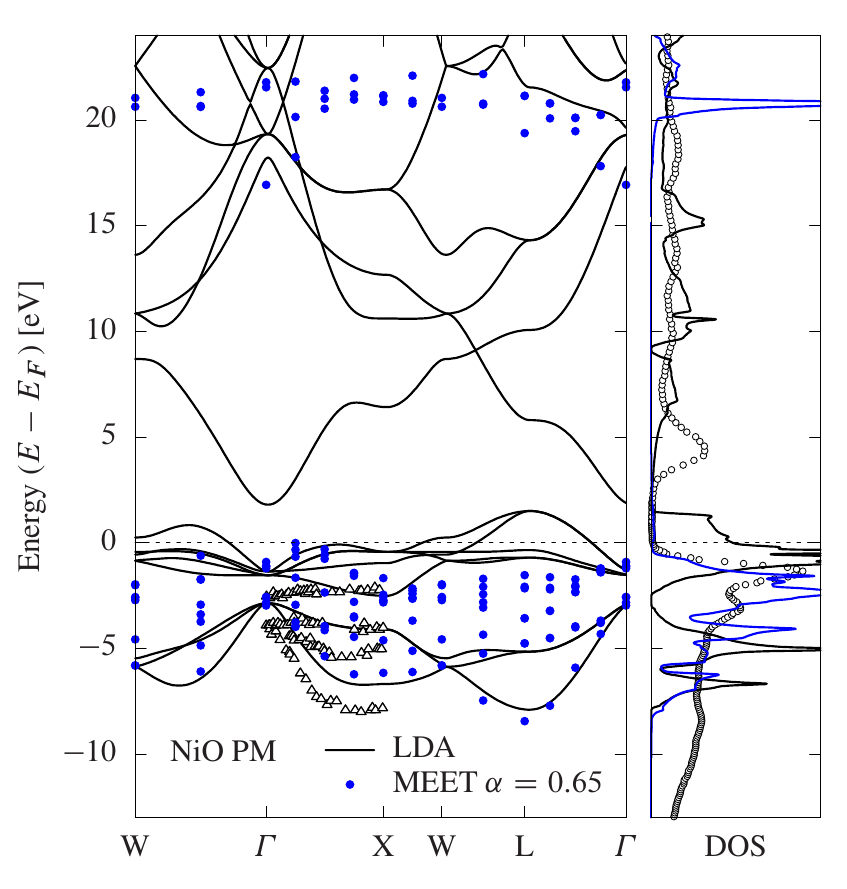}
\caption{Band structure and DOS of the paramagnetic phase of bulk NiO: experiment (dots, triangles) \textit{vs} LDA (black solid line) and MEET at $\delta^{(1)}$ level (blue dots). Note that for the MEET we only reported the energies associated with occupation numbers larger than 0.20, which are the same already present in LDA, in order to not charge the figure; this does not change the conclusions. 
The experimental band structure is taken from Ref.~[\onlinecite{PhysRevB.44.3604}]. Note that the experimental band structure and DOS are taken from different experiments; for sake of comparison we align the high-energy valence band along the $\Gamma-X$ directon with the high-energy peak in the occupied part of the DOS.
}
\label{Fig:NiO_band-structure}
\end{center}
\end{figure}

In Fig.\ (\ref{figurePM}) we report the calculated and measured\cite{PhysRevB.44.1530,PhysRevLett.39.1229,PhysRevLett.53.2339,PhysRevB.54.10245,PhysRevB.44.6090} PES of paramagnetic MnO, FeO, CoO, and NiO, together with  the occupation numbers that play a role into the spectrum for the reported energy range. Note that the results of NiO have already been reported in Ref. [\onlinecite{stefano}], we show them again here for sake of comparing the PES across the whole series of oxides studied in this work. The trend observed for NiO is also found for the other metal oxides, namely the opening of a band gap, albeit a largely overestimated gap. 
As we will discuss in more detail below this is not obvious. While for NiO the picture concerning the occupation numbers is similar to that of the atomic limit of the Hubbard dimer at one-half filling for which our method is exact to all orders \cite{stefano}, this is not the
case for the other materials, for which the occupation numbers show a different behavior.
As discussed in Ref. [\onlinecite{stefano}], the overestimation of the band gap is expected using the lowest approximation to $\delta^{R/A}(\omega)$ and higher approximations such as $\delta^{R/A,(2)}(\omega)$ are expected to close the band gap. Work to include higher order corrections in an effective way is in progress. We find that at the level of $\delta^{R/A,(1)}$ the MEET tends to underestimate the band widths with respect to LDA and deforms the bands around the Fermi level with the creation of extra poles, which leads to a band gap opening. In particular we observe that the MEET tends to collapse the $d$-like conduction bands around some high-energy value, which is responsible for the large overestimation of the band gap. This is illustrated for NiO in Fig. (\ref{Fig:NiO_band-structure}).

\begin{figure}[t]
\begin{center}
\includegraphics[width=0.45\textwidth]{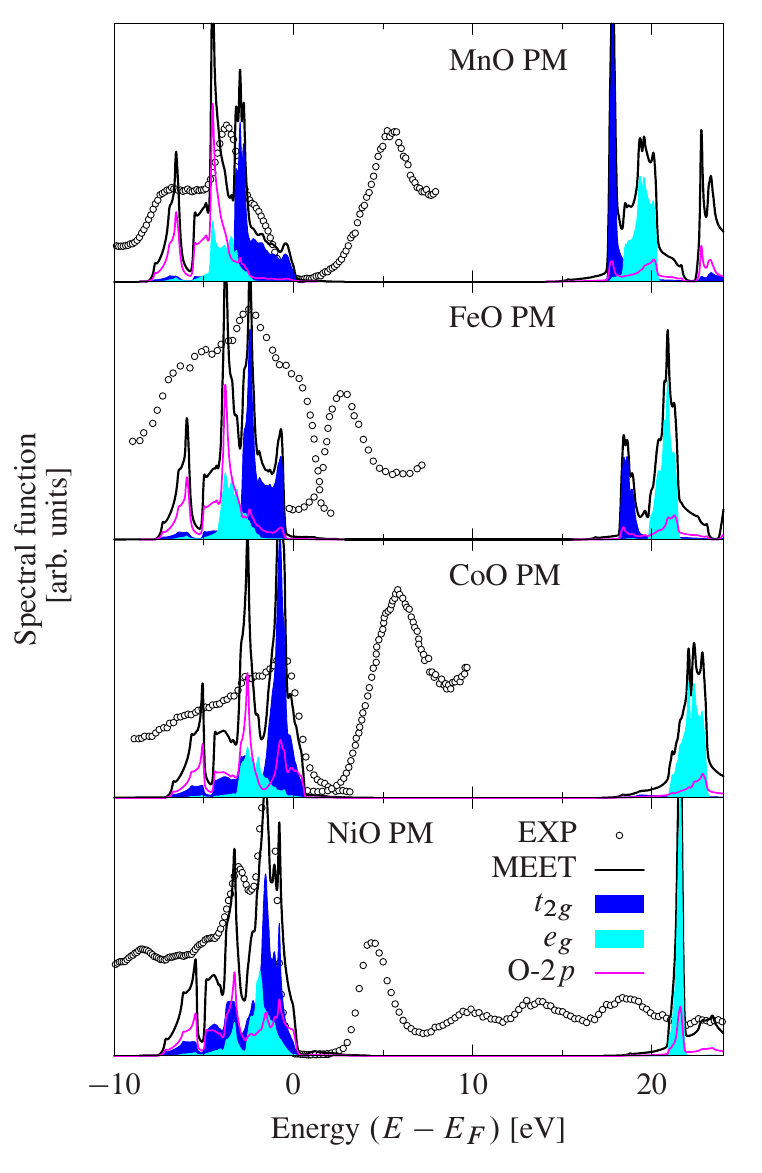}
\caption{Contributions of the transition metal $d$ states of $t_{2g}$ (blue filled) and $e_g$ (light blue filled) symmetry and of the oxygen $2p$ state (magenta solid line) to the total
spectral function (black solid line) of MnO, FeO, CoO, and NiO in the paramagnetic phase.} 
\label{figureProj}
\end{center}
\end{figure}

The analysis of the occupation numbers in case of the PM phase reveals that all the four oxides are characterized by fractional occupation numbers in the band gap region, but whereas NiO has occupation numbers clustered around 0.5 (which reassembles the same scenario of the atomic limit of the Hubbard dimer at  half-filling \cite{stefano}), going from MnO, through FeO, to CoO, the occupation numbers slowly move towards 1 or 0, with CoO showing a similar picture as for the AF phase.\cite{stefano} This trend can be interpreted as follows. The TM in NiO is a $d^8$, with six electrons in the $t_{2g}$ orbitals and two in the $e_g$. The two electrons in the $e_g$ show a similar physics as the atomic limit of the Hubbard dimer at one-half filling, which occupation numbers equal 0.5. The TM in CoO is a $d^7$, with the $t_{2g}$ filled and one electron in the $e_g$ (which is also in line with the projected DOS in Fig.\ (\ref{figureProj})), which behaves as the atomic limit of the Hubbard dimer at one-forth filling, which occupation numbers equal either 0 or 1. MnO and FeO show a more complex scenario, since Mn and Fe have five and six electrons in the $d$ orbitals, respectively, with partially filled $t_{2g}$ and $e_g$  orbitals;  this is a highly degenerate situation with many occupation numbers that are fractional.  

In Fig.\ (\ref{figureProj}) we report the projection of the spectral function on the oxigen $2p$ and metal $e_g$ and $t_{2g}$ orbitals. The composition is in general agreement with DMFT-based calculations on these systems found in literature.\cite{Nekrasov2013,Dyachenko2012,2017arXiv170502387Z} It essentially reflects the filling of the $d$ bands moving from Mn to Ni: the $t_{2g}$ and $e_g$ are initially split between the upper valence bands and the lower conduction bands with the $t_{2g}$ contribution to the latter decreasing along the series until disappearing in NiO, in which the upper conduction bands have mainly $e_g$ character. The order of the $t_{2g}$ and $e_g$ is not completely in agreement with DMFT calculations in literature, but this is also due to the use of the lowest approximations to $\delta^{R/A}(\omega)$. Higher order approximations can change this picture. 


\section{Conclusions and outlooks \label{Sec:Conclusions}}
In this work we apply our recently derived many-body effective energy theory for the calculation of photoemission spectra to typical strongly-correlated systems, namely MnO, FeO, CoO, and NiO in the paramagnetic phase. All these systems have partially filled $d$ states and are described as metals in standard approximations. We show that our theory gives a qualitative good description of the photoemission spectrum of these systems. Besides NiO, which shows a very similar scenario as observed in the atomic limit of the Hubbard dimer at one-half filling, for which our method is exact at all order of approximations, our theory performs well also for the other TMOs, which deviates from this scenario as far as the occupation numbers are concerned.  The band gap, however, is strongly overestimated; this is traced back to the use of the lowest order approximation to the theory, which depends only on the one-body and two-body reduced density matrices. The next level of approximation, which depends also on the three-body reduced density matrix, is expected to close the gap; however higher-order approximations can quickly become uncontrolled and yield unphysical results, such as negative spectral functions. Various steps can be taken to improve the theory, in particular one could i) look for alternative ways to truncate the series, which guarantee well-known sum rules to be fulfilled at all orders ; ii) re-sum higher order terms in order to work only with effective one and two-body reduced density matrices. Work in these directions is ongoing.

\begin{acknowledgments}
This work is supported by ANR (project no. ANR-18-CE30-0025-01).
 \end{acknowledgments}

\bibliography{bibliography}

\end{document}